\documentclass[twocolumn,aps,pra]{revtex4}
\usepackage{epsfig}
\usepackage[english]{babel}
\usepackage{latexsym}
\usepackage{graphics}
\usepackage{subfigure}
\usepackage{epsfig}
\usepackage{graphicx}
\usepackage{dcolumn}
\usepackage{amsmath}
\usepackage{hyperref}
\usepackage{amssymb}

\begin{document}

\title{Tracing origins of asymmetric momentum distribution for polar molecules in strong linearly-polarized laser fields}
\author{S. Wang$^{1,2}$, J. Y. Che$^{1}$, C. Chen$^{1}$, G. G. Xin$^{3,4}$, and Y. J. Chen$^{1,*}$}

\date{\today}

\begin{abstract}

We study the ionization dynamics of oriented HeH$^+$ in strong linearly-polarized laser fields
by numerically solving the  time-dependent Schr\"{o}dinger equation.
The calculated photoelectron momentum distributions  for parallel orientation show a striking asymmetric structure. 
With a developed model pertinent to polar molecules, we trace the electron  motion in real time. 
We show that this asymmetric structure arises from the interplay of the Coulomb effect and the permanent dipole in strong laser fields.
This structure can be used to probe the degree of orientation  which is important in ultrafast experiments for polar molecules. we also check our results for other polar molecules such as CO and BF.

\end{abstract}
\affiliation{1.College of Physics and Information Technology, Shaan'xi Normal University,Xi'an710119, China\\
2.School of Physics, Hebei Normal University, Shijiazhuang 050024, China\\
3.School of Physics, Northwest University, Xi'an 710127, China\\
4.Shaanxi Key Laboratory for Theoretical Physics Frontiers, Northwest University, Xi'an710069, China} 

\maketitle

\section{Introduction}
Above threshold ionization (ATI) is a basic process in
strong-laser-matter interactions \cite{Kulander1,Lewenstein2,Becker2002},
which has wide applications in attosecond science \cite{Eckle,Blaga,Paul2018}.
One of the important observables in ATI is the photoelectron momentum distribution (PMD),
which encodes rich dynamical information of ATI and the structural information of the target \cite{Okunishi2008,Micheau2009}.
Many experimental and theoretical efforts including numerical  solution of the time-dependent Schr\"{o}dinger equation (TDSE) and analytical treatments based on strong-field approximations (SFA) \cite{Lewenstein2,Becker2002} have been devoted to this issue.
It is shown that for the simple case of an atom  interacting with a linearly-polarized single-color  laser field,
the PMD is symmetric with respect to the axis perpendicular to the laser polarization \cite{Huismans2011}.
The phenomenon also holds for a symmetric linear molecule aligned parallel or
perpendicular to the laser polarization \cite{Meckel2014,Liu2016}.
This symmetry is easily understood with considering the time-domain symmetry of the laser pulse and the symmetric geometry of the target.
The situation is different for an asymmetric molecule even it is oriented along the laser polarization.
For example, numerical studies have shown that the PMD for HeH$^{2+}$ is not symmetric \cite{Bandrauk2014}.
Naturally, one can think that this asymmetry, a typical characteristic of PMD from polar molecules,
is a result of the asymmetric geometry of the target. 
However, the underlying physical mechanism is unclear.

Here, we study ATI from oriented HeH$^+$ \cite{Banyard,Wiesemeyer} with a permanent dipole (PD) numerically and analytically.
The HeH$^+$ molecule (the first product of chemical reaction in the cosmos) has a stable ground state
which can be manipulated in present experiments \cite{Wustelt,Yue}.
The TDSE is first solved with Born-Oppenheimer (BO) approximations, then extended to  non-BO cases.

\begin{figure}[t]
\begin{center}
{\includegraphics[width=8.5cm,height=6.3cm]{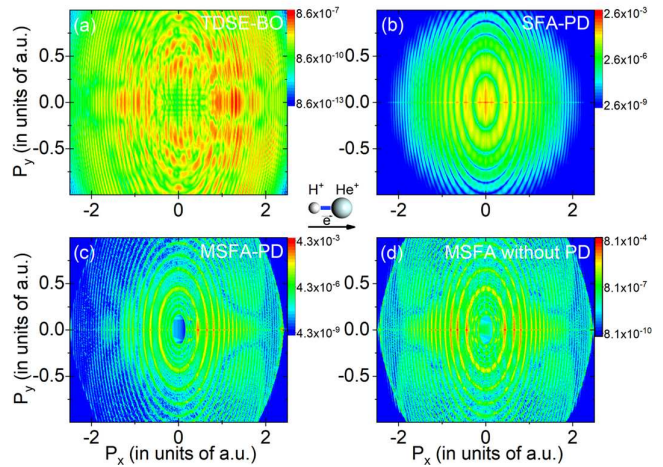}}
\caption{Photoelectron momentum distributions of HeH$^+$ obtained with different methods at $\theta=0^{\circ} $, internuclear distances are fixed at $R=1.4 $a.u.. (a): TDSE-BO;
(b): SFA-PD; (c): MSFA-PD; (d): MSFA without considering  PD.
The insets indicate  the positions of these two nuclei He and H and the unit vector $\vec{\mathbf{e}}$ along the laser polarization. The  parameters used here are $I=1.5\times10^{15}$W/cm$^{2}$, $\omega_0=0.114$ a.u. ($\lambda=400$ nm)
}
\label{fig1}
\end{center}
\end{figure}

The TDSE predictions of PMD for HeH$^+$ show a striking asymmetry.
This asymmetry, however, can not be described by the  SFA with considering the PD effect (SFA-PD).
With further considering the Coulomb modification in SFA-PD (MSFA-PD),
this asymmetry is reproduced in our simulations. By contrast, this asymmetry also disappears
when neglecting the PD effect in MSFA treatments (see Fig. \ref{fig1}).
The comparisons provide deep insights into the origin of the asymmetric PMD.
We show that the interplay of the Coulomb effect and the PD effect is mainly responsible for this asymmetric structure.
When the PD effect destroys the symmetry of the ionization in two consecutive half-cycles,
the Coulomb effect further destroys the symmetry of the ionization in the half laser cycle where the ionization mainly occurs. These complex dynamic processes are memorized by photoelectrons, resulting in the asymmetry of PMD.

\section{Numerical method}
In the BO case, the Hamiltonian of  the HeH$^+$ system studied here
has the form of H$(t)=H_0+\mathbf{r}\cdot \mathbf{E}(t)$ (in atomic units of $\hbar= e = m_{e} = 1$).
Here, the term $H_0=\mathbf{p}^2/2+V(\mathbf{r})$ is the field-free Hamiltonian, and  $V(\mathbf{r})=-\sum_{j=1,2}\frac{Z(R,|\mathbf{r}-\mathbf{R}_j|)}{\sqrt{\xi+|\mathbf{r}-\mathbf{R}_j|^2}}$  is the  Coulomb potential with
$Z(R,r_j)=Z_{ji}\exp[-\rho(R){r}_j^2]+Z_{jo}$,
 $r_j=|\mathbf{r}-\mathbf{R}_j|=\sqrt{(x-x_j)^2+(y-y_j)^2}$ and $j=1,2$.
Here, $Z_{1}$ and $Z_{2}$ are the effective  charges
for the He  and  H centers, respectively. The indices i and o
denote the inner and outer limits of $Z_{1}$ and $Z_{2}$.
$\mathbf{R}_{1}$ and $\mathbf{R}_{2}$ are the positions of the He and H
nuclei that have the coordinates of $(x_1 ,y_1)$ and $(x_2 ,y_2)$ in
the $xy$ plane, with $x_{1/2}=\pm R_{1/2}\cos\theta$, $y_{1/2}=\pm
R_{1/2}\sin\theta$, $R_1=M_{H}R/(M_{He}+M_{H})$ and $R_2=M_{He}R/(M_{He}+M_{H})$.
$M_{He}$ and $M_{H}$ are  masses of He and H nuclei and $\theta$ is the orientation angle which is the angle between the molecular axis and the laser polarization.
For HeH$^+$, we have used the  parameters of  $Z_{1i}=2/3,Z_{2i}=1/3,Z_{1o}=4/3,$ and $Z_{2o}=2/3$.
$\xi=0.5$  is the softening  parameter.
$\rho(R)$ is the screening parameter, which is adjusted such
that the ionization potential $I_p(R)$ of the model
molecule at the distance R matches the real one.
For example, for the equilibrium separation of $R=1.4$ a.u. with $I_p=1.65$ a.u., $\rho(R)=0.94$.
and for the stretched case of $R=1.8$ a.u. with $I_p=1.5$ a.u., $\rho(R)=1.21$.
For non-BO cases, we follow the procedure introduced in detail in \cite{Li}.

The term $\mathbf{E}(t)$ is the electric field  which has the form of
$\mathbf{E}(t)=\vec{\mathbf{e}}{E}_x(t)$
with ${E}_x(t)=f(t)E_{0}\sin{(\omega_{0}t)}$.
Here, $\vec{\mathbf{e}}$   is the unit vector along the laser polarization which is along the $x$  axis, 
 ${E}_0$ is the maximal laser amplitude relating to the peak intensity $I$ of  ${E}_x(t)$,  
$\omega_{0}$ is the laser frequency  and $f(t)$ is the envelope function.
We use trapezoidally shaped laser pulses with a total duration of 8 optical cycles and linear ramps of two optical cycles.
The details for solving  TDSE of $i\dot{\Psi}(t)=$H$(t)\Psi(t)$ with spectral method \cite{Feit}
and obtaining the PMD can be found in \cite{Gao,Shang}.
Unless mentioned elsewhere, the  parameters used here are $I=1.5\times10^{15}$W/cm$^{2}$, $\omega_0=0.114$ a.u. ($\lambda=400$ nm),
$R=1.4$ a.u. and $\theta=0^{\circ} $ at which 
the He (H) nucleus is located on the right (left) side.

To analytically study the ATI of polar molecules, as in \cite{Madsen2010}, we first incorporate the PD effect into SFA. With the saddle-point approximation, the tunneling amplitude for the  photoelectron with the drift momentum $\mathbf{p}$ and the complex ionization time $t_s=t_0+it_x$ can be written as $F(\mathbf{p},t_s)\equiv F(\mathbf{p},t_0)\propto\sum_s\big[\beta\textbf{E}(t_s)\cdot \textbf{d}_i(\textbf{p}+\textbf{A}(t_s))e^{-iS}\big]$ with $S\equiv S(\textbf{p},t_s)$ and $\beta\equiv({1/det(t_s)})^{1/2}$. Here, $\textbf{d}_i(\textbf{v})=\langle \textbf{v}| \hat{\textbf{r}}|0\rangle$ is the bound-continuum transition matrix element, $S(\textbf{p},t')=\int_{t'}\{(\textbf{p}+\textbf{A}(t''))^{2}/2+[I_{p}+\mathbf{E}(t'')\cdot \mathbf{D}]\}dt''$ is the quasiclassical action. $\textbf{A}(t)$ is the vector potential of the electric field $\textbf{E}(t)$ and $\textbf{D}$ is the permanent dipole.
The value of $\textbf{D}$ used in the expression of $F(\mathbf{p},t_s)$ has the following form: $\mathbf{D}=\mathbf{D}_{0}=\left\langle 0|\hat{\mathbf{r}}|0\right\rangle -\sum_{\alpha=1}^{\alpha=2} Z_{\alpha}\mathbf{R}_{\alpha}$. Here, the term of $\left\langle 0|\hat{\mathbf{r}}|0\right\rangle$ is the ground-ground transition matrix element, $\mathbf{R}_{1}$ ($\mathbf{R}_{2}$) is the vector from the origin of coordinates to the nucleus He (H) whose effective charges are $Z_{1}=2/3$ ($Z_{2}=1/3$).
In this paper, we will call the above SFA with considering the PD effect the SFA-PD. In our TDSE cases, for $R=1.4$ a.u., $D=-0.13$ a.u. and for  $R=1.8$ a.u., $D=-0.27$ a.u.. These values will also be used in our model treatments. 
We mention that with using the expression of $\textbf{D}=\textbf{D}_{0}-\textbf{D}_{1}$ where 
$\textbf{D}_{1}=\left\langle 1|\hat{\mathbf{r}}|1\right\rangle -\sum_{\alpha=1}^{\alpha=2} Z_{\alpha}\mathbf{R}_{\alpha}$, 
which approximately considers the strong coupling between the ground state and the first excited state of the asymmetric system \cite{Shang}, 
the results obtained are in qualitative agreement with those obtained with $\textbf{D}=\textbf{D}_{0}$.

Then as in \cite{Brabec,Goreslavski}, we solve the Newton equation
$\ddot{\mathbf{r}}(\mathbf{p},t)=-\mathbf{E}(t)-\nabla_\mathbf{r} V(\mathbf{r})$
for each SFA electron trajectory ($\mathbf{p}$, $t_s$), with  initial conditions \cite{Beckeroribit} $\dot{\mathbf{r}}(\mathbf{p},t_0)=\mathbf{p}+\mathbf{A}(t_0)$
(the exiting momentum) and
$\mathbf{r}(\mathbf{p},t_0)=Re(\int^{t_0}_{t_s}[\mathbf{p}+\mathbf{A}(t')]dt')$ (the exiting position).
Here,  the real part $t_0$ of $t_s$  is considered as the exiting time.
The final Coulomb-modulated  drift momentum is obtained with $\mathbf{p}_f=\dot{\mathbf{r}}(\mathbf{p},t\rightarrow\infty)$,
which is related to the amplitude $F(\mathbf{p},t_0)$.
In this paper, we will call the above Coulomb-modified SFA-PD  the MSFA-PD.
According to the SFA, the exiting time $t_0$  agrees with the ionization time $t_i$
at which the  instantaneous
energy $E_a(t)=[\dot{\mathbf{r}}(\mathbf{p},t)]^2/2+V(\mathbf{r})$ becomes larger than zero.
However, as discussed in \cite{chen2019},  the MSFA
predicts a time lag $t_{d}=t_i-t_0$ with $t_d>0$.
The influence of this lag on the dynamics of  laser-driven system is remarkable.
It can change the direction of the drift momentum, e.g., from $p_x<0$ to  $p_x>0$ in our cases,
and importantly increases the contributions of long trajectory to ATI.
As a result, the PMD of SFA is modified remarkably.
Below, we will show that for  HeH$^+$, the time lag $t_d$
which is related to electrons exiting the potential along the H side plays an important role in the PMD of the polar molecule.

\section{Asymmetric PMD}
In Fig. \ref{fig1}(a), we show the TDSE photoelectron momentum distribution for HeH$^+$ in BO cases. This distribution shows a remarkable asymmetry with respect to
the axis of ${p}_x=0$. The SFA-PD result, as shown in Fig. \ref{fig1}(b), however, shows a symmetric distribution,
implying that the PD effect itself does not induce this asymmetry.
With considering the Coulomb effect, as seen in Fig. \ref{fig1}(c), the MSFA-PD clearly reproduces this asymmetry. In addition,
the interference structures of the distribution in Fig. \ref{fig1}(c) are also comparable to the TDSE ones in Fig. \ref{fig1}(a).
By contrast, when assuming $D\equiv0$ in the MSFA-PD simulations, a symmetric distribution                                                                                                                                                                                                                                                                                                                                                                                                                                                                                                                                                                                              is observed, as seen in Fig. \ref{fig1}(d).
This result implies that the asymmetric Coulomb potential itself also does not induce the asymmetric distribution.
From these comparisons, one can conclude that the asymmetric PMD of polar molecules is closely related to
the interaction of PD and Coulomb effects.
\begin{figure}[t]
\begin{center}
{\includegraphics[width=8.5cm,height=6.6cm]{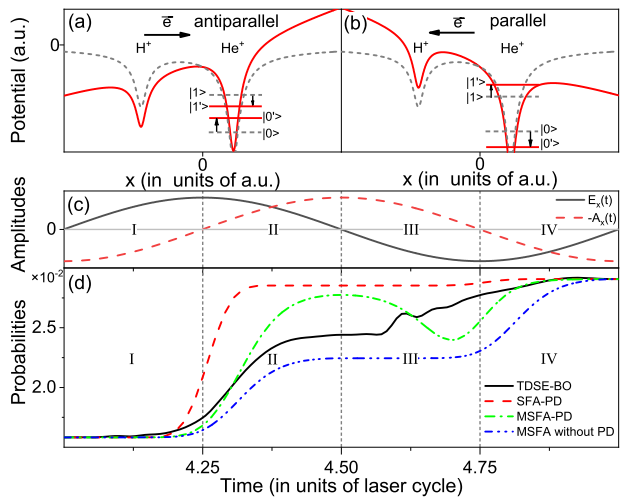}}
\caption{A sketch of the PD effect and the Coulomb-induced ionization time lag.
(a) and (b): the field-free (dotted curves) and laser-dressed (solid curves) asymmetric potentials and these  two field-free
($|0\rangle$ and $|1\rangle$) and
laser-dressed ($|0'\rangle$ and $|1'\rangle$) lowest states   when the laser polarization is antiparallel (a)
or parallel (b) to the permanent dipole (which is directing from the He nucleus to the H nucleus); (c): the fields of $E_x(t)$ and -$A_x(t)$;
(d): time-dependent ionization probabilities calculated with  TDSE-BO,
SFA-PD, MSFA-PD and MSFA without considering PD for $I=1.5\times10^{15}$W/cm$^{2}$, $\lambda=400$ nm,
$R=1.4$ a.u..
For comparison,  each model curve in (d) is multiplied vertically by a constant factor to match the TDSE one.
The vertical lines divide the laser cycle into four parts (I-IV) as shown.
}
\label{fig2}
\end{center}
\end{figure}

\section{Potential mechanism}
To understand the potential mechanism, first, we introduce the PD effect on ATI of polar molecules.
In Figs. 2(a) to 2(c), we present a sketch of the laser-dressed states related to the PD effect and  the electric field $E(t)$ in one laser cycle.
As discussed in \cite{Shang}, the PD effect depends strongly on the polarization of the laser field.
For the first half cycle with $E(t)>0$ at which  electrons escape from the H side, due to the PD effect, the energy of the ground state $|0\rangle$ (the first excited state $|1\rangle$) is dressed up (down), making the ionization easier to occur (see Fig. \ref{fig2}(a)).
For the second half cycle with $E(t)<0$ at which  electrons tunnel out of the potential along the He side, the situation reverses and the ionization is more difficult to occur (see Fig. \ref{fig2}(b)). This phenomenon has been termed asymmetric ionization of polar molecules \cite{Bandrauk2}.
The classical prediction \cite{Schafer1,Corkum} of drift momenta $p_x(t)=-A_x(t)$ is also presented in Fig. \ref{fig2}(c),
indicating that the electron born in the time regions I and IV (II and III) contributes to $p_x<0$ ($p_x>0$).

With the introduction of the PD effect, in Fig. \ref{fig2}(d), we further compare time-dependent
ionization probabilities in one laser cycle predicted with different methods.
The TDSE ones are obtained with evaluating $I(t)=1-\sum_{m=1}^{m=15}|\langle m|\Psi(t)\rangle|^2$. Here, $|m\rangle$ is the bound eigenstate of H$_0$ obtained through imaginary-time propagation. The model ones are obtained with calculating $I(t)=\sum_{\mathbf{p},t}|F(\mathbf{p},t_0)|^2$ at $E_a(t)>0$. 
The SFA-PD results give a good illumination on the asymmetric ionization,
with showing a remarkable increase just around the time of $t=4.25T$  in the first half cycle
and a small increase just around $t=4.75T$ in the second half cycle.
Here, $T=2\pi/\omega_0$.
The TDSE results, however, show a larger increase around a time obviously later than $t=4.25T$  (about 100 attoseconds) in the first half cycle and
a smaller increase around a time near $t=4.75T$ for the second half cycle.
These phenomena of asymmetric ionization and striking ionization time lag in two consecutive half-cycles are reproduced by the MSFA-PD.
Considering that the ionization mainly occurs in the first half of the cycle, we mainly focus on the ionization time lag during the first half of laser cycle. Here, the value of the ionization time lag is taken as the time difference between the time when the electric field rises to its peak value and the instant around which the ionization probability increases remarkably. 
Without considering the PD effect, the MSFA fails for predicting the asymmetric-ionization phenomenon.
These results in Fig. \ref{fig2}(d) are in agreement with those in Fig. \ref{fig1}.
From Fig. \ref{fig2}, this asymmetry in PMD can be easily understood.
For SFA-PD and MSFA without PD, the contributions of regions II and III versus regions I and IV to ionization are almost the same.
For TDSE and MSFA-PD, the contributions of regions II and III are remarkably larger than those of regions I and IV.
These results indicate that due to both the PD effect which induces the dominating ionization for electrons exiting the potential along the H side,
and the Coulomb effect which induces the large ionization time lag related to the H-side ionization,
the PMD for $p_x>0$ has larger amplitudes than  that for $p_x<0$,
explaining the asymmetry in PMD observed in Figs. 1(a) and 1(c).
This remaining difference between TDSE and MSFA-PD predictions, especially for the second half cycle may be due the omission of the excited-state effect \cite{Shang} in MSFA-PD treatments, which is not easy to incorporate into the MSFA at present.

\begin{figure}[t]
\begin{center}
{\includegraphics[width=8.5cm,height=5cm]{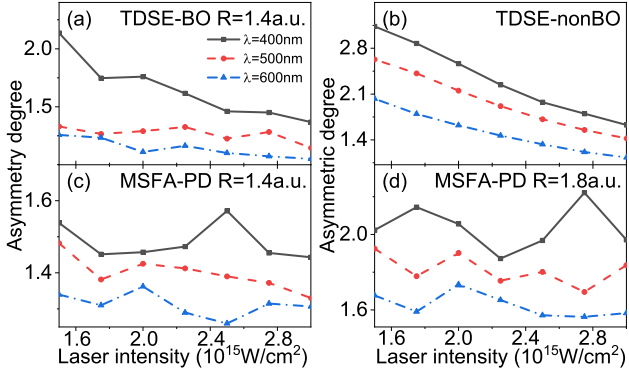}}
\caption{Asymmetry degrees evaluated with different methods for different laser parameters at $\theta=0^{\circ}$. (a):  TDSE-BO with $R=1.4$ a.u.;
(b): TDSE-nonBO; (c): MSFA-PD with $R=1.4$ a.u.;
(d): MSFA-PD with $R=1.8$ a.u..}
\label{fig3}
\end{center}
\end{figure}

\section{Asymmetry degree}
We have also extended our simulations to other laser parameters and non-BO cases. Relevant results are presented in Fig. \ref{fig3},
where we plot the asymmetry degree of the PMD which is defined as the ratio of the total amplitudes with $p_x>0$ to those of $p_x<0$.
The TDSE results with $R=1.4$ a.u. in Fig. \ref{fig3}(a) show that on the whole, this degree is larger for shorter laser wavelengthes and lower laser intensities. These parameter-dependent phenomena are basically reproduced by the MSFA-PD in Fig. \ref{fig3}(c).
When the nuclear motion is considered, the non-BO TDSE results in Fig. \ref{fig3}(b) are similar to the BO ones in Fig. \ref{fig3}(a),
but the  curves are more smoothing and the evaluated degrees are somewhat larger than the corresponding BO ones.
Due to the PD effect, the HeH$^+$ molecule usually stretches rapidly towards larger $R$ in strong laser fields \cite{Li}.
For the present laser parameters, the maximal stretching distance is around $R=1.8$ a.u..
For comparison, MSFA-PD simulations at $R=1.8$ a.u. are also presented in Fig. \ref{fig3}(d), these parameter-dependent phenomena are similar to those in Fig. \ref{fig3}(b). 

\begin{figure}[t]
\begin{center}
{\includegraphics[width=8.5cm,height=5cm]{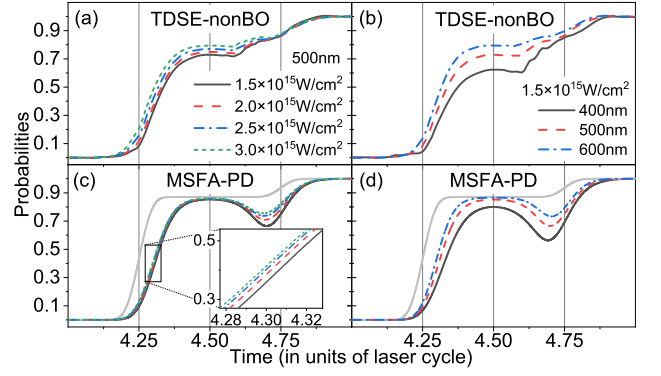}}
\caption{Calculated time-dependent ionization probabilities of HeH$^+$ with various laser parameters for TDSE-nonBO (upper panels), and for MSFA-PD with $R=1.8$ a.u. (lower panels).  For comparison, we normalize the ionization probabilities within one laser cycle. In (a) and (c), we show the ionization probabilities of $\lambda=500$ nm with different laser intensities. In (b) and (d), we show the ionization probabilities of $I=1.5\times10^{15}$W/cm$^{2}$ with different laser wavelengths. In (c), the inset shows the enlarged results of the corresponding time-dependent 
ionization probabilities. The gray lines in (c) and (d) are calculated by SFA-PD 
at $I=1.5\times10^{15}$W/cm$^{2}$ and $\lambda=500$ nm for comparison (see the context for details).
}
\label{fig4}
\end{center}
\end{figure}

As discussed above, the asymmetric structure arises from the interplay of effects of the Coulomb potential  
and the permanent dipole in strong laser fields. Hence, the asymmetry degree of PMD reflects the 
ionization dynamics of polar molecules.  
In Fig. \ref{fig4}, we show the ionization probabilities calculated by TDSE-nonBO (upper panels)
and MSFA-PD with $R=1.8$ a.u. (lower panels) for different laser parameters. 
Combining the results of Fig. \ref{fig3}(b) and  Figs. \ref{fig4}(a) and \ref{fig4}(b), one can observe that
the ionization time lag as well as the asymmetry degree of the PMD  decreases with the increase of laser intensity 
at the same laser wavelength or decreases with the increase of laser wavelength at the same laser intensity. 
These phenomena of parameter-dependant ionization time lag are reproduced by the MSFA-PD in Figs. \ref{fig4}(c) and \ref{fig4}(d),
indicating the applicability of this model. 
We mention that the normalized ionization probabilities 
within one laser cycle obtained by SFA-PD without including the coulomb effect for different laser parameters are almost the same, 
so we only show the results of SFA-PD at one certain laser parameter here, as shown  by the gray lines
in Figs. \ref{fig4}(c) and \ref{fig4}(d).
Comparisons between the results of MSFA-PD and SFA-PD in Figs. \ref{fig4}(c) and \ref{fig4}(d) also indicate that the Coulomb effect is mainly responsible for the parameter-dependant ionization time lag of polar molecules in strong laser fields.

\section{Potential applications}
\begin{figure}[t]
\begin{center}
{\includegraphics[width=8.5cm,height=6cm]{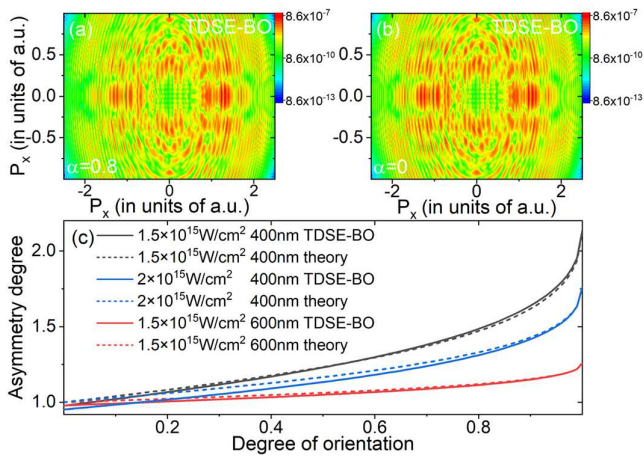}}
\caption{Photoelectron momentum distributions of HeH$^+$  obtained with  TDSE-BO at different degrees of orientation $\alpha$. (a): $\alpha=0.8$;
(b): $\alpha=0$. In (c), we show the asymmetry degree
evaluated with TDSE (solid) and a simple formula (dashed) at different $\alpha$  and laser parameters.
}
\label{fig5}
\end{center}
\end{figure}

Our further research shows that the asymmetric structure also depends on the molecular orientation. 
Compared to the case of $\theta=0^{\circ}$, the He (H) nucleus is located on the left (right) side for $\theta=180^{\circ}$. With this inversion of the molecular geometry, our results indicate the inversion of the asymmetric structure in PMD for both numerical and analytical simulations. One of the potential applications of the asymmetric PMD is
to evaluate the degree of orientation $\alpha=(n_u-n_d)/(n_u+n_d)$ achieved in experiments \cite{Frumker,chen2013}. Here, $n_u$ ($n_d$) is the number of the polar molecule pointing up (down).
In Fig. \ref{fig5}, we show the PMDs of TDSE-BO at different degrees of orientation $\alpha$ with assuming perfect alignment.
One can observe that with the decrease of the value of $\alpha$, this asymmetry in PMD diminishes and for $\alpha=0$ (random orientation),
a symmetric PMD is observed. In other words, this asymmetric structure is sensitive to the degree of orientation.
Using this phenomenon, one can evaluate the value of $\alpha$. In Fig. \ref{fig5}(c), we plot the asymmetry degree versus $\alpha$,
obtained with TDSE-BO simulations and a simple formula:
\begin{equation}
\gamma(\alpha,\beta)=\frac{\left(\sqrt{\frac{1+\alpha}{2}}+\sqrt{\frac{1-\alpha}{2\beta}}\right)^2}{\left(\sqrt{\frac{1+\alpha}{2\beta}}+\sqrt{\frac{1-\alpha}{2}}\right)^2}.
\label{eq1}
\end{equation}
Here, $\gamma$ is the asymmetry degree at
$\alpha$, and $\beta$ is that at the perfect orientation obtained with TDSE simulations.
This formula can be easily deduced with the assumption that for each $|p_x|$,
the PMD of $p_x>0$ differs from that of  $p_x<0$ only for a constant factor $\gamma$.
The TDSE and analytical results agree with each other,
implying  that this formula can be used to evaluate the degree of orientation.
\begin{figure}[t]
\begin{center}
{\includegraphics[width=8.5cm,height=6cm]{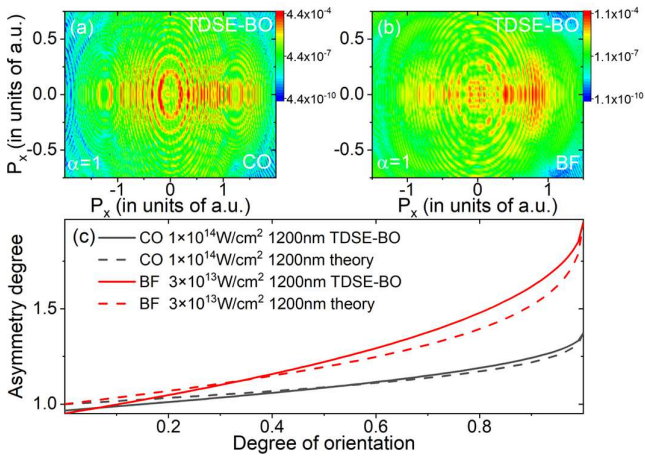}}
\caption{Photoelectron momentum distributions of CO and BF  obtained with  TDSE-BO at $\alpha=1$ (perfect orientation).
In (c), we show the corresponding asymmetry degree
evaluated with TDSE (solid) and a simple formula (dashed) at different $\alpha$  for these two polar molecules. The laser parameters used are as shown.}
\label{fig6}
\end{center}
\end{figure}

We have also extended our TDSE simulations to other polar molecules such as CO and BF with using the model potentials as introduced in \cite{Shang}. For CO with $I_p=0.52$ a.u., we have used the laser parameters of $I=1\times10^{14}$W/cm$^{2}$ and $\lambda=1200$ nm, and for BF with a smaller ionization potential of $I_p=0.41$ a.u., we have used the parameters of $I=4\times10^{13}$W/cm$^{2}$ and $\lambda=1200$ nm. Relevant results are presented in Fig. \ref{fig6}. In Figs. \ref{fig6}(a) and \ref{fig6}(b), the PMDs for these two polar molecules at perfect orientation also show the striking asymmetry, similar to the case of HeH$^+$ in Fig. \ref{fig1}.

In particular, this asymmetry for BF with a larger PD is more remarkable  than that for CO. For these cases, the simple formula of $\gamma(\alpha,\beta)$ introduced in Fig. \ref{fig5} also gives a good description for the dependence of the asymmetry degree on the degree of orientation, as seen in Fig. \ref{fig6}(c).

\section{Conclusion}
In summary, we have studied ATI of HeH$^+$ in strong linearly-polarized single-color laser fields. The PMD of HeH$^+$ shows
a striking asymmetry.   
With the development of a model which considers
both the Coulomb effect and the PD effect, we show that
this remarkable asymmetry arises from attosecond  dynamics of polar molecules in strong laser fields.
The PD effect induces a preferred ionization when the active electron escapes from a certain side of the molecule.
The Coulomb effect further induces a large time lag (about 100 attosecond) between tunneling and ionization,
which remarkably changes the symmetry of PMD.
This mechanism is essential for polar molecules with a large PD such as CO and is believed to has important influences on
other strong-field processes such as high-order ATI and non-sequential double ionization.
Besides monitoring the electron tunneling dynamics with attosecond resolution,
the asymmetry of PMD for polar molecules can also be used for calibrating the degree of orientation.

\section*{ACKNOWLEDGMENTS}
This work is supported by the National Natural Science Foundation of China (Grant No. 91750111),
the National Key Research and Development Program of China (Grant No. 2018YFB0504400),
and the Fundamental Research Funds for the Central Universities, China (Grant No. GK201801009).

\end{document}